\documentclass[aps,groupedaddress,showpacs,twocolumn]{revtex4}

\usepackage{amsmath}
\usepackage{graphicx} 
\newcommand{\pzhetp}{{\it Pis'ma Zh. \'{E}ksp. Teor. Fiz. }}
\newcommand{\jetpl}{{\it JETP Lett. }}

\begin{document}   
\pagestyle{empty}
\title{Two dimensional modulational instability in photorefractive media}
\author{ M. Saffman$^1$,  Glen McCarthy$^2$, and  Wieslaw Kr\'olikowski$^2$ }

\address{ 1)   Department of Physics,
University of Wisconsin, 1150 University Avenue,  Madison, Wisconsin 53706, USA\\
2) CUDOS (Centre for Ultrahigh bandwidth Devices and Optical Systems) and 
 Laser Physics Centre, Research School of Physical Sciences and Engineering, 
Australian National University, Canberra ACT 0200, Australia}
 \date{\today}
\begin{abstract} 
We study theoretically and experimentally the modulational instability of broad
optical beams in photorefractive nonlinear media. We demonstrate the impact of  
the anisotropy of the nonlinearity on the growth rate of periodic 
perturbations. Our findings are confirmed by  experimental measurements in a strontium 
barium niobate photorefractive crystal.
\end{abstract}
\pacs{42.65.Hw, 42.65.Jx}
\maketitle

\vspace{.5cm}

\pagestyle{empty}

A plane wave propagating in a medium with focusing nonlinearity is unstable with 
respect to the generation of  small scale filaments \cite{BesTal}. This so called 
modulational instability (MI) phenomenon, has been extensively studied because of 
its  importance as a factor limiting the propagation of high power beams. 
Filamentation may also be identified as the first stage in the development of 
turbulent fluctuations in the transverse profile of a laser beam\cite{zak92}. In 
addition, MI is often considered as a precursor for the formation of spatial and/or temporal  optical solitons. As far as  optics is concerned, MI has been studied in media with various mechanisms of nonlinear response including cubic \cite{BesTal}, quadratic \cite{MI_SHG}, nonlocal \cite{nonlocal1,nonlocal2}  and inertial \cite{MI_slow,MI_incoherent} types of nonlinearity.   
Importantly, MI  is not restricted to nonlinear optics but has 
 also been studied in many other nonlinear systems  including 
fluids \cite{fluids}, plasmas \cite{plasma} and matter waves \cite{BEC}. 

In the 
context of optical beam 
propagation in nonlinear materials MI has usually been considered in media with 
spatially isotropic nonlinear properties. Recently a great deal of theoretical and 
experimental efforts have been devoted to studies of nonlinear optical effects and 
soliton formation in photorefractive crystals  \cite{PR1,PR2,PR3}. While these media 
exhibit strong nonlinearity at very low optical power  their 
nonlinear response is inherently anisotropic \cite{zapra}. The anisotropy 
causes a number of observable effects including astigmatic self-focusing of optical beams \cite{astigmatism}, elliptically shaped solitary solutions\cite{zams}, geometry-sensitive interactions of solitons \cite{ks}, and fixed optical pattern orientation \cite{mamsaflinear}. 

Several previous studies of MI in the context of photorefractive 
media were  limited to a 
1-dimensional geometry  where the anisotropy is absent \cite{MI_PR1,MI_PR2,MI_PR3},
 and  the physics is similar to 
the standard saturable nonlinearity \cite{saturable}. On the other hand, in a real 
physical situation  where one   deals with finite sized beams,  the  anisotropic 
aspects of  the photorefractive nonlinear response are expected  to play a 
significant role. Some previous work \cite{msz1,zamspra2} already indicated the importance of anisotropy in the transversal break-up of broad beams propagating in biased photorefractive crystals. However, no detailed analysis of this phenomenon was carried out. In this paper we study the MI of optical beams in  photorefractive media taking into account the full 2-dimensional anisotropic model  of the photorefractive nonlinearity.

Time independent propagation of an optical beam ${\mathcal E} (r,z)=(A/2)e^{\imath (k z- \omega t)}+ c.c.$ in a nonlinear medium with a weakly varying index of refraction is 
governed by the parabolic equation 
%
\begin{equation}
\frac{\partial A}{\partial  z} - \frac{i}{2 k}  \nabla_\perp ^{2} A(r,z) =  
 i\: k \frac{ n_2(r,z)}{n_0} A(r,z) \;.
\label{parabolic}
\end{equation}
Here $r=(x,y)$ and $z$ are transverse and axial coordinates,  
$\nabla_\perp = \hat{x} (\partial /\partial x) + \hat{y} (\partial / \partial y),$
$k= 2 \pi n_0/\lambda,$ $\lambda$ is the wavelength in vacuum, $\omega = 2 \pi c/\lambda,$ $c$ is the speed of light, and $n=n_0+ n_2(r,z)$ is the refractive index, with $n_0$ the spatially uniform background index, and $n_2$ the spatially varying nonlinear increment.

In the case of a photorefractive screening nonlinearity the optical beam propagates through  a
photorefractive crystal externally biased with a DC electric field. The beam 
excites charges which after migrating due to diffusion and drift in the  applied 
field, are subsequently trapped by impurity or defect centers. The effective  
nonlinearity (refractive index change)  is proportional 
to the low frequency electric field $\bf E$ created by light induced charge redistribution \cite{PR1,PR2,zapra}. In the situation of interest here where the
optical field is linearly polarized along $\hat x$ which coincides with the  crystalline $\hat c$ axis
the  nonlinear increment to the refractive index is given by $n_2(r,z)=-\frac{1}{2}n_0^3r_{33} E_x(r,z),$ with $r_{33}$ the relevant component of the electro-optic tensor, and $E_x$ the $\hat x$ component of the low frequency electric field in the medium. 

It is convenient to describe the nonlinear material response in terms of the quasi-static potential induced by the   optical field. As shown in the appendix the resulting set of dimensionless equations is 
\begin{eqnarray}
&&\frac{\partial A}{\partial  z} - i \nabla_\perp ^{2} A =  
 i\: \frac{\partial \phi}{\partial x} A,
\label{par2}
\\
 &&\tau \frac{\partial}{\partial t}\left[\nabla_\perp\cdot\left( \hat \epsilon_n \nabla_\perp\phi \right) \right]+
\nabla_\perp^2\phi+\nabla_\perp\phi\cdot\nabla_\perp\ln\bar I
\nonumber\\
&&= \frac{\partial}{\partial x}\ln\bar I
+\frac{E_{\rm ph}}{E_{\rm ext} \bar I}\frac{\partial |A|^2}{\partial x}
+ \alpha \frac{\nabla_\perp^2 \bar I}{\bar I}
\label{potential}
\end{eqnarray}
where $\bar I = (1+|A|^2)/[1- \xi \nabla_\perp\cdot(\hat \epsilon_n \nabla_\perp \phi)].$
The coordinates and variables have been normalized using the scalings given in the appendix with the addition of 
$|A|^2/\tilde I\rightarrow |A|^2$ with $\tilde I=2 I_s/(\epsilon_0 n_0 c).$

Equation (\ref{potential}) describes the most general situation when the 
electrostatic potential in the crystal is induced by two distinct transport 
mechanisms: drift of charges  in the biasing DC field plus photogalvanic field and their diffusion. 
The relative strength of the diffusion and drift terms is determined by the dimensionless parameter 
$$
\alpha=\frac{k_BT }{eE_{\rm ext}l_\perp}.
$$
The diffusion contribution which leads to spatially asymmetric stimulated scattering dominates at large transverse wavenumbers of order $k_D.$ On the other hand the drift terms give the dominant contribution to the spatially symmetric MI which is prominent at much smaller transverse wavenumbers. Thus  the term proportional to $\alpha$ in Eq. (\ref{potential}) is often  neglected when studying MI.

The initial linear stage of the filamentation instability 
may be investigated by putting 
\begin{equation}    
A(r,z)=A_0e^{\imath  \beta z}\left(1+a e^{\Gamma z + \imath {\bf q}\cdot {\bf r} 
+\imath\Omega t}
+ b e^{\Gamma^* z - \imath {\bf q}\cdot {\bf r}-\imath\Omega t}  \right), 
\label{ansatz}
\end{equation}
where $\Gamma$ represents the growth rate of the perturbation characterized by the 
transverse wavevector  $\bf q$,  and frequency detuning   $\Omega$.

The steady state solution to Eq. (\ref{potential}) in the one dimensional plane wave limit with
$\alpha=\xi=0$  is 
$\partial \phi/\partial x =  (1+E_{\rm ph}/E_{\rm ext})|A_0|^2/(1+|A_0|^2)$ where $A_0$ is the amplitude of the plane wave which is assumed to vanish for $x\rightarrow\pm\infty$. It is therefore convenient to renormalize the potential as $\phi= (1+E_{\rm ph}/E_{\rm ext})|A_0|^2 x/(1+|A_0|^2)+\tilde \phi$ so that the equations of motion become
\begin{eqnarray}
&&\frac{\partial A}{\partial  z} - i  \nabla_\perp ^{2} A =  
 i\:  \left(\frac{ (1+E_{\rm ph}/E_{\rm ext})|A_0|^2}{1+|A_0|^2}+ \frac{\partial \tilde\phi}{\partial x}\right) A \;,
\label{pwe2}\\
&&\tau \frac{\partial}{\partial t}\left[\nabla_\perp\cdot\left( \hat \epsilon_n \nabla_\perp\tilde\phi \right) \right]+
\nabla_\perp^2\tilde\phi+\nabla_\perp\tilde\phi\cdot\nabla_\perp\ln\bar I
\nonumber\\
&&=\frac{1-(E_{\rm ph}/E_{\rm ext})|A_0|^2}{1+|A_0|^2} \frac{\partial}{\partial x}\ln\bar I
+
\frac{E_{\rm ph}}{E_{\rm ext} \bar I}\frac{\partial |A|^2}{\partial x}
+ \alpha \frac{\nabla_\perp^2 \bar I}{\bar I}\nonumber\\
\label{potential2}
\end{eqnarray}
with $\bar I = (1+|A|^2)/[1- \xi \nabla_\perp\cdot(\hat \epsilon_n \nabla_\perp \tilde\phi)].$
The ansatz (\ref{ansatz})  then gives $\beta=(1+E_{\rm ph}/E_{\rm ext})|A_0|^2/(1+|A_0|^2)$ which provides a continuous transition to the 1D solution.

\begin{figure}[!h]
\begin{center}
\includegraphics[width=.4\textwidth]{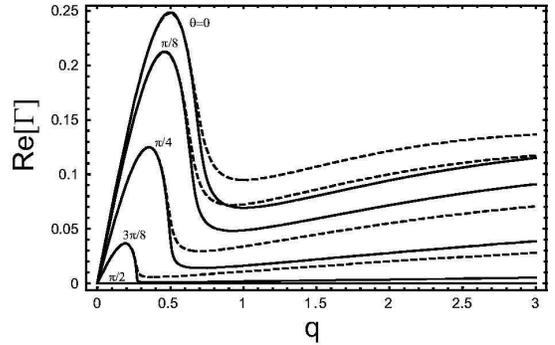}
\end{center}
\caption{Growth rate as a function of spatial frequency $q$ for few values of the  transverse angle $\theta$ for $A_0=1,$ $\Omega=0,$ $E_{\rm ph}=0,$ 
$\epsilon_y/\epsilon_c=0.53,$ 
and
$\xi= 0.42.$ The solid lines show the diffusionless case  ($\alpha=0)$ and the dashed lines correspond to $\alpha=0.046.$ } 
\label{angularspectrum}\end{figure}

Solving the linear problem defined by Eqs. (\ref{ansatz}-\ref{potential2})
gives the dispersion relation
\begin{widetext}
\begin{equation}
\Gamma^2= q^2 \left[ \frac{2  |A_0|^2}{ 1+|A_0|^2}
\frac{\frac{\cos^2(\theta)}{1+|A_0|^2}\left(1+\frac{E_{\rm ph}}{E_{\rm ext}}\right)+i\alpha q\cos(\theta)}
{1+\hat \epsilon_n(\theta)\left[i\Omega \tau -i\frac{q\cos(\theta)\xi}{1+|A_0|^2}\left(1-|A_0|^2\frac{E_{\rm ph}}{E_{\rm ext} } \right)+ \alpha \xi q^2\right] }
 -q^2 \right]
\label{hpr}
\end{equation}
\end{widetext}
where $\hat\epsilon_n(\theta)=\cos^2(\theta)+\frac{\epsilon_y}{\epsilon_c}\sin^2(\theta)$ with $\epsilon_y$ the static dielectric tensor component along $\hat y$ and $\theta$ the angle of $\bf q$ with respect to the $\hat x$ axis.  
In the limit of a single transverse dimension ($q_y=0$)  without diffusion
Eq.(\ref{hpr})
 reduces to the formula for the growth rate in saturable nonlinear 
media \cite{MI_slow,saturable}.

The instability growth rate is given by ${\rm Re}[\Gamma]=\sqrt{{\rm Re}[\Gamma^2]+ |\Gamma^2|}/\sqrt{2}$. In Fig.\ref{angularspectrum} we show the growth rate as a function of the 
spatial frequency $q=|\bf q|$ for a few values of  the angle $\theta$
using parameters characteristic of a photorefractive crystal as given in the appendix.   The growth  rate depends strongly on the angular orientation of the initial 
perturbation. In particular, it always attains the largest  value when the wave 
vector of the perturbation coincides with the direction of the applied electric 
field ($\theta=0$). As $\theta$ departs from zero the amplification of the 
perturbation decreases, and the growth rate becomes less strongly peaked at small $q$. The growth rate is an even function of $q$ provided $\Omega=0$ and inspection of Eq. (\ref{hpr}) shows that it falls of for large $q$ as $1/q.$ Interestingly  Eq.(\ref{hpr}) predicts there is no  
instability for perturbations with wave vectors perpendicular to the direction of 
the applied field.

\begin{figure}[!h]
\begin{center}
\includegraphics[width=.4\textwidth]{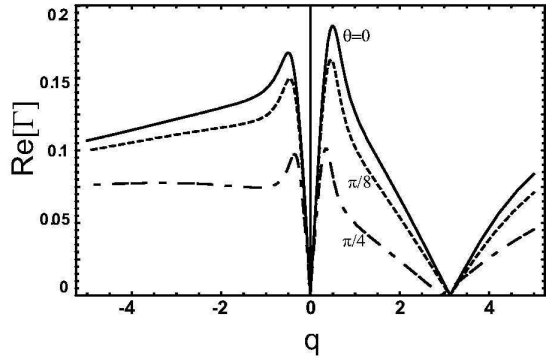}
\end{center}
\caption{Growth rate of frequency shifted perturbations vs spatial frequency $q$ for few values of the of transverse angle $\theta$ for $A_0=1,$ $\Omega\tau=1,$ $E_{\rm ph}=0,$ 
$\epsilon_y/\epsilon_c=0.53,$
$\alpha=0.046,$  
and
$\xi= 0.42.$  } 
\label{angularspectrum2}\end{figure}

When the perturbation is frequency shifted with respect to the plane wave  $(\Omega \ne 0)$ the growth rate becomes an asymmetric function of $\bf q.$ This is shown in Fig. \ref{angularspectrum2} for $\Omega\tau=1.$ Positive $q$ in the figure corresponds to the direction of a plane wave that experiences two-wave mixing gain. We see that the growth rate has several maxima as a function of $q$ in the presence of a frequency shift.

Finally we note that the angular dependence takes on a simple form in the small $q$ limit where the diffusion contribution can be neglected. 
Putting $\alpha=\xi=0$ in (\ref{hpr}) and assuming no frequency shift so $\Omega=0$ we obtain
\begin{equation}
{\rm Re}[\Gamma]=q\left[ \frac{2|A_0|^2}{(1+|A_0|^2)^2}\left(1+\frac{E_{\rm ph}}{E_{\rm ext}} \right)\cos^2\theta-q^2\right]^{1/2}.
\label{simplerate}
\end{equation}
This latter expression shows clearly the decline of the instability growth rate with increasing angle away from the $\hat x$ axis.


\begin{figure}[!h]
\begin{center}
\includegraphics[width=.4\textwidth]{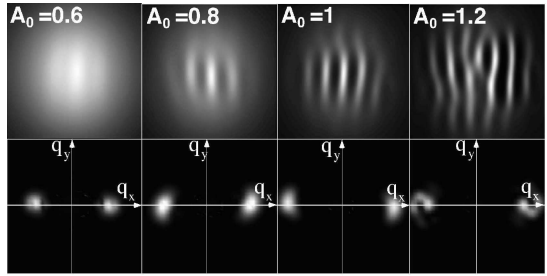}
\end{center}
\caption{Intensity distribution (top) and its spatial spectrum (bottom)of a
gaussian beam with initial random noise, after propagation over a distance of 5mm 
in a photorefractive crystal. The size of the computational window is $200\mu \times 200\mu$. A zero-frequency component has been removed from the spectrum.}
\label{noise_calc}\end{figure}


\begin{figure}[!h]
\begin{center}
\includegraphics[width=.4\textwidth]{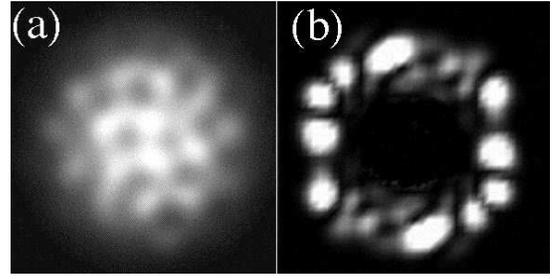}
\end{center}
\caption{Intensity distribution (a) and its spatial spectrum (b) of a
broad Gaussian beam with a superimposed initial random noise, after propagation over a distance of 5mm in an isotropic nonlinear medium. For  clarity of presentation the zero-frequency component has been removed from the spectrum.}
\label{isotropic}\end{figure}
\begin{figure}[!h]
\begin{center}
\includegraphics[width=.4\textwidth]{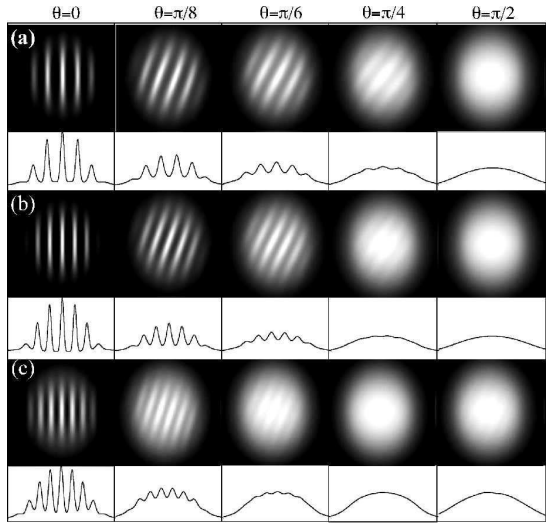}
\end{center}
\caption{Intensity distribution of the optical  beam in a biased photorefractive 
crystal with an initially imposed periodic perturbation as a function of the 
angular orientation of the perturbation and its wave vector ($q$).  Peak intensity $A_0^2$=1; DC biasing electric field applied along horizontal ($x$) direction. The size of the computational window is $200\mu \times 200\mu$.} \label{periodic_calc}
\end{figure}

In order to verify the reliability of the  linear approximation discussed above we 
resorted to numerical analysis of the full 2-dimensional model governing  
propagation of optical beams in a photorefractive medium Eqs. (\ref{par2},\ref{potential}).  
For simplicity and comparison with the experimental conditions discussed below we only considered the frequency degenerate ($\Omega=0$) and short Debye length ($\alpha=\xi=0$) limit. We also assumed the lack of a photogalvanic effect ($E_{ph}=0$).
We used a split  step Fast Fourier  transform code  to solve the propagation equation (\ref{par2}) and a finite 
difference technique to find  the electrostatic potential and refractive index 
change induced by the beam. Results of these calculations are shown in Fig.(\ref{noise_calc}-\ref{periodic_calc}).  In all cases the direction of the applied DC field is horizontal (along the x-axis).
In Fig.(\ref{noise_calc})  we show a few examples of numerical simulations depicting results of 
propagation of the wide Gaussian beam with initial random perturbation of its 
amplitude, for a few values of the beams peak intensity. These graphs illustrate the 
inherently anisotropic nature of the instability. Initially random perturbations 
lead to amplification of the perturbation with almost zero y-component of the 
wave vector. This leads to appearance of beam modulation in the form of roughly vertically  oriented stripes. 
Graphs in the bottom row display the spatial spectrum (a zero frequency 
component has been removed for clarity of presentation) of the resulting 
intensity distribution. Notice that the spatial frequency with the highest growth rate varies with the peak intensity of the beam, which is in agreement with the 
prediction of Eq.  (\ref{simplerate}). For comparison Fig.\ref{isotropic} shows the intensity 
distribution obtained with the Gaussian beam propagating in a standard isotropic 
saturable medium. The beam experiences modulational instability but this 
time all amplified spatial frequencies are located on a ring  reflecting the isotropic 
nature of the nonlinear process.  

The complete absence of instability for $\theta=\pi/2$ and its apparent one-dimensional character as depicted in Fig.(\ref{noise_calc}) are a direct consequence of the anisotropy of the nonlinear response of the photorefractive medium. The light induced focusing power is roughly  three times stronger in the direction of the applied DC field than in the direction perpendicular to it \cite{astigmatism}. Unlike the isotropic system where all spatial frequencies corresponding to highest growth rate are amplified (Fig.\ref{isotropic}), in the photorefractive  crystal the highest gain is experienced only by perturbations with $q_y\approx 0$ ($\theta\approx 0)$. Therefore only these frequencies will  contribute to the initial stages of the modulational instability described by the linear theory. Spatial perturbations  with nonzero $q_y$ components have much weaker growth rates and will play an important role only after the 1D structure with $q_y \approx 0$ has reached  sufficiently high intensity\cite{mamsaflinear}. Then the full 2-dimensional break-up and subsequent filamentation of the beam will follow \cite{msz1,MI_stripe2}.  However, the full analysis of such a process is beyond the scope of the present paper.  

Next we simulated propagation of a broad  Gaussian beam with its amplitude  
perturbed by a spatially periodic modulation.
 The angle $\theta$ which determines the 
angular orientation of the perturbation with respect to the direction of applied 
DC field was varied from $\theta=0$ to $\theta=90^\circ$. The strength of the 
perturbation (relative to the peak intensity of the beam) was less than $10^{-2}$. Results of the propagation of this beam over 
a distance of five millimetres  are shown in Fig.\ref{periodic_calc}. Each row of this figure 
corresponds to a different spatial frequency of the initial perturbation. A decrease in the amplification of the perturbation with increasing angle  
$\theta$ is evident.

\begin{figure}[!h]
\begin{center}
\includegraphics[width=.4\textwidth]{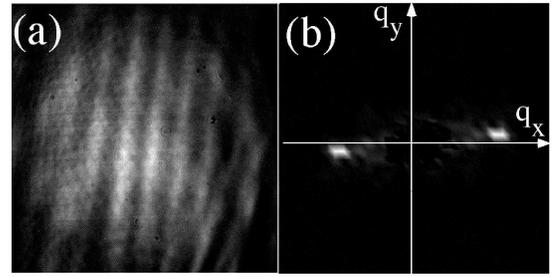}
\end{center}
\caption{Experimentally observed self-induced modulational instability of the 
initially Gaussian beam (width=200$\mu$). (a) light intensity distribution; 
(b) spatial spectrum (with zero frequency component removed). Amplitude of the 
external biasing DC field $E_{ext}=3kV/cm$. DC field applied in horizontal direction. The width of the experimental window is 320$\mu \times 320\mu$ }
\label{exp1}\end{figure}

\begin{figure}[!h]
\begin{center}
\includegraphics[width=.4\textwidth]{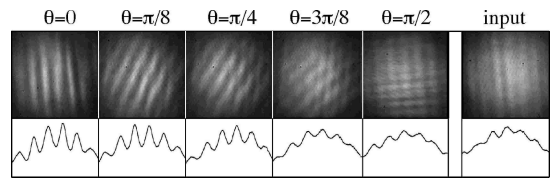}
\end{center}
\caption{Experimentally observed  dependence of the modulational instability of 
the broad  Gaussian beam on the angular orientation of initial periodic 
perturbation. Amplitude of the external  DC electric field $E_{ext}=2.2kV/cm$ Field applied along the  horizontal direction.}\label{exp2}
\end{figure}

To verify our theoretical findings we conducted experiments using a crystal of 
photorefractive strontium barium niobate as the nonlinear medium. The 
experimental setup is analogous to that used in our earlier studies of 
photorefractive soliton formation \cite{PR3}. The crystal was 5x5x10 mm in size with the optical beam propagating along
the 10 mm axis and a DC electric field of 1.1 kV applied along the 5 mm long
$\hat c$-axis. 
The optical beam (1mW) from a solid state laser ($\lambda=532~\rm nm$) was loosely 
focused at the input face of the crystal. The output intensity distribution was 
imaged by a CCD camera and stored in a computer. The crystal was illuminated by a 
broad white light beam which was used to control the degree of saturation. 
Typically, the peak intensity of the incident beam was of the same order as the average intensity of the white light background.  We used either an unperturbed beam or beam 
with superimposed weak periodic perturbations. Results of the experiments are shown 
in Fig.\ref{exp1}-\ref{exp2}. Fig.\ref{exp1}(a) shows the light intensity distribution at the exit 
facet of the crystal (after 10 mm of propagation) in the  case where the incoming beam was 
not intentionally perturbed.  It is evident that nonlinearity induces modulational 
instability which leads to the formation of quasi 1-dimensional vertical stripes oriented 
perpendicularly to the direction of the applied DC field. Figure \ref{exp1} (b) depicts the corresponding Fourier spectrum of the outgoing beam where the two distinct peaks with almost zero $q_y$ components clearly indicate the anisotropic character of the instability. The presence of the small y-component in the spectrum is the result of a slight misalignment of the crystal.  
 

\begin{figure}[!h]
\begin{center}
\includegraphics[width=.4\textwidth]{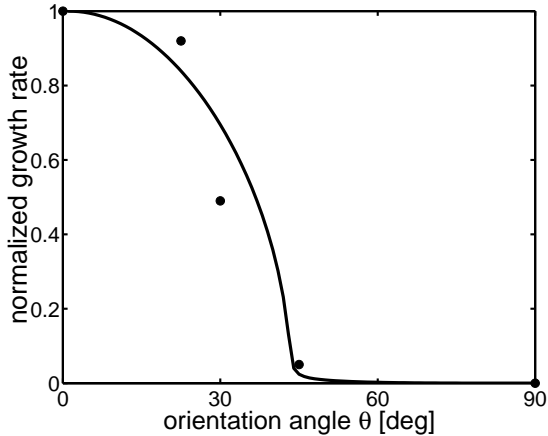}
\end{center}
\caption{Experimentally measured  normalized growth rate of an initial periodic 
perturbation imposed onto the broad Gaussian beam as a function of an angular orientation  of the initial pattern. Dots - experimental points; line - theoretical fit (formula (\ref{simplerate}))}
\label{exp3} 
\end{figure}

Finally  we investigated the role of anisotropy in modulational instability of a broad beam with an initially imposed periodic perturbation. To this end the incoming Gaussian beam was initially transmitted through a parallel plate which resulted in the appearance of a weak spatial periodic modulation of the beam wavefront. By rotating the plate we were able to change the orientation of this modulation. The perturbed beam subsequently propagated through the biased photorefractive crystal.  Results of this experiment are shown in Fig.\ref{exp2}. 
Grey scale plots in the top row of this figure represent the light 
intensity distribution at the  output face of the photorefractive crystal corresponding  to different  angular orientations of the 
periodic pattern characterized by the angle ($\theta$).  Graphs in the bottom row illustrate the corresponding intensity profile. 
As   Fig.(\ref{exp2}) clearly shows the amplification of the perturbation decreases rapidly as the angle  departs from $\theta=0$.  The rightmost plot shows the intensity pattern at the input face of the crystal.  For better visualisation we plot  in Fig. \ref{exp3}  the experimentally  measured growth rate (normalized to its maximum value)  as a function of the angle $\theta$.  The points represent experimental data while the line is a theoretical fit (Eq.(\ref{simplerate}) with $E_{ph}=0$,  $A_0=2.6$, $\xi=0.42$, $\alpha=0$ and $q=0.35$. Again, the drop in amplification of the perturbation for increasing $\theta$  is evident.


In conclusion, we investigated modulational instability of plane waves and finite 
beams in photorefractive nonlinear media biased with a DC electric field. We showed that the growth rate of 
perturbation is affected by the inherent anisotropy of the nonlinear 
response. It is  highest for perturbations whose wave-vectors correspond to the 
direction of the biasing DC field. For arbitrarily oriented perturbations the effect 
of anisotropy manifests itself  in a decrease of  the effective strength of the  
nonlinear response until it reaches zero for wave-vectors perpendicular to the 
direction of the field. Our theoretical predictions were confirmed by experimental 
observations in strontium barium niobate crystals.

{\bf Acknowledgement}

The work of W.K and G.M. has been supported by the Australian Research Council.

\appendix

\section{Derivation of equations}

The set of equations describing the optical properties of a photorefractive crystal,  known as the Kukhtarev equations\cite{kuk}, are

\begin{subequations}
\begin{eqnarray}
&&\frac{\partial N_D^+}{\partial t}
=(\beta + \sigma I_{em})(N_D-N_D^+)-\gamma_r n_e N_D^+ \label{eq.nd}\\
&&\rho=e(N_D^+-N_A-n_e)\\
&&{\bf J}=e\mu n_e {\bf E} + \mu k_B T\nabla n_e + \beta_{ph}(N_D-N_D^+)I_{em}\hat c \nonumber \\
\\
&&\nabla\cdot (\epsilon_0 \hat\epsilon {\bf E})=\rho \label{eq.gauss}\\
&&\frac{\partial\rho}{\partial t}+\nabla\cdot{\bf J}=0.
\end{eqnarray}
\end{subequations}
Here  $N_D, N_D^+, N_A,$ and $n_e$ are the density  
of donors, ionized donors, acceptors, and conduction electrons, 
$\beta$ and $\sigma$ are the coefficients of thermal and photoexcitation, $I_{em}$ is the optical intensity, $\gamma_r$ is the electron recombination coefficient, $-e$ is the charge on an electron, 
$\epsilon_0$ is the permeability of vacuum, $\hat\epsilon$ is the static dielectric tensor, $k_B$ is the Boltzmann constant, $T$ the temperature, $\mu$ the electron mobility, $\rho$ the charge density, $\bf J$ the current, and $\bf E$ the static electric field.  Note that the coefficient $\sigma$ includes the photogalvanic contribution due to $\beta_{ph}$ so that we could write
the total photoexcitation coefficient as  $\sigma = \sigma_1+\beta_{ph}.$

We analyze these equations following the approach of  Ref. \cite{zapra}.  In the absence of thermal or photoexcitation $n_e=0$ so the condition $<N_D^+>=<N_A>,$ where $<>$ denotes a spatial average, ensures bulk charge neutrality. The
negatively charged acceptors do not participate in the photoexcitation dynamics so the density $N_A$ is fixed and serves to limit the magnitude of the photoexcited space charge field.  
To analyze the Kukhtarev equations we assume $N_D\gg N_A\gg n_e.$
Gauss's law (\ref{eq.gauss}) then gives 
$$
N\equiv \frac{N_D^+}{N_A}=1+\frac{1}{eN_A}\nabla\cdot\epsilon_0\hat \epsilon \bf E.
$$
Introducing the Debye wavenumber for charge motion along the $\hat c$ axis as $k_D=e\sqrt{N_A/(k_B T \epsilon_0 \epsilon_c)},$
where $\epsilon_c$ is the component of the dielectric tensor along $\hat c$ and the characteristic field $\tilde E= k_B T k_D/e=
e N_A/(\epsilon_0 \epsilon_c k_D)$ we can write the last expression as 
$$
N=1+\frac{1}{k_D \tilde E}\nabla\cdot\hat \epsilon_n \bf E.
$$
with $\epsilon_n$ the static dielectric tensor divided by $\epsilon_c.$

The assumption of fast carrier recombination implies that $\partial n_e/\partial t$ can be set to zero in the equation for charge continuity. It follows that 
\begin{eqnarray}
&&\frac{\partial M}{\partial t}+ 
\frac{1}{\epsilon_0 \epsilon_c k_D \tilde E} \nabla\cdot\left[e\mu n_e {\bf E} + \mu k_B T\nabla n_e \right. \nonumber\\
&&~~~~~~\left.  +\beta_{ph}(N_D-N_D^+)I_{em}\hat c\right]=0,
\label{eq.time1}
\end{eqnarray}
 where $M=N-1.$
To proceed we use Eq. (\ref{eq.nd}) to write
$$
n_e=-\frac{1}{\gamma_r N}\frac{\partial N}{\partial t}
+\frac{\beta}{\gamma_r}(1+I_{em}/I_s)\frac{N_D/N_A-N}{N}
$$
where $I_{s}=\beta/\sigma$ is the saturation intensity for which  
the rate of thermal excitation equals the rate of photoexcitation.
 With fast carrier recombination and $N_D\gg N_A$ we have
$$
n_e\simeq \frac{\beta}{\gamma_r N}(1+I)\frac{N_D}{N_A}
$$
with $I=I_{em}/I_s.$ 
The characteristic density of electrons associated with $I_s$ is 
$n_0=(\beta/\gamma_r)(N_D/N_A)$ so the electron density can be written as
\begin{equation}
n_e=n_0 \frac{1+I}{1+M}.
\end{equation}
Furthermore the photogalvanic term can be written as
\begin{eqnarray}
\beta_{ph}(N_D-N_D^+)I_{em}&\simeq&
\beta_{ph}N_DI_{em}\nonumber\\
&=&\beta_{ph}\frac{N_A n_e \gamma_r N }{\sigma}\frac{I}{1+I}\nonumber\\
&=&  \beta_{ph}\frac{N_A n_0 \gamma_r  }{\sigma}I.
\end{eqnarray}

Defining the characteristic relaxation time of the electric field $t_0=
\frac{\epsilon_0\epsilon_c}{e \mu n_0}$ and the photogalvanic field 
$E_{ph}=\frac{ \beta_{ph} \gamma_r N_A  }{e\mu \sigma}$ Eq. (\ref{eq.time1}) can be written as  
\begin{eqnarray}
&&t_0 \frac{\partial M}{\partial t}+ 
\frac{1}{ k_D \tilde E} \nabla\cdot\left[
 \frac{1+I}{1+M}{\bf E} + \frac{ \tilde E}{k_D}\nabla \frac{1+I}{1+M} 
 +  E_{ph} I\hat c\right]=0.\nonumber\\
\label{eq.time3}
\end{eqnarray}
 This equation coincides with Ref. \cite{zapra}, Eq. (2) with $\chi=0$ and $\delta=0.$ 

We are interested in the situation where the optical beam is  small compared to the size of the nonlinear medium.  
The externally applied bias field is $E_{\rm ext}=V/L_x$ with $V$ the applied voltage and $L_x$ the width of the medium along $\hat x$ which is taken to coincide with the $\hat c$ axis.  It is convenient to subtract this field from the optically induced field  so that the field ${\bf E}_{r}=  {\bf E}- E_{\rm ext}\hat x$ vanishes at the boundaries of the medium.
Using  ${\bf E}_{r}$ instead of ${\bf E}$ in Eq. (\ref{parabolic}) 
results in only a small change in the wavenumber of the beam
which has no physical importance for this work. We then introduce a potential through the relation ${\bf E}_{r}=-\tilde E \nabla_\perp \phi$ so that Eq. (\ref{eq.time3}) can be written as
\begin{eqnarray}
& & t_0 \frac{\partial}{\partial t}\left[\nabla\cdot\left( \hat \epsilon_n \nabla\phi \right) \right]+\frac{1+I}{1+M}
\nabla^2\phi+\nabla\phi\cdot\nabla\frac{1+I}{1+M}
\nonumber\\
&&= \frac{E_{\rm ext}}{\tilde E}\frac{\partial}{\partial x}\frac{1+I}{1+M}
+\frac{E_{\rm ph}}{\tilde E}\frac{\partial I}{\partial x}
+ \frac{1}{k_D} \nabla^2 \frac{1+I}{1+M}.
\label{app_pot1}
\end{eqnarray}

In the situation of interest here where the
optical field is linearly polarized along $\hat x$ 
the  nonlinear increment to the refractive index is given by $n_2(r,z)=-\frac{1}{2}n_0^3r_{33} E_x(r,z),$ with $r_{33}$ the relevant component of the electro-optic tensor, and $E_x$ the $\hat x$ component of the low frequency electric field in the medium. 
In a paraxial approximation the optical field therefore satisfies the parabolic equation
\begin{equation}
\frac{\partial A}{\partial  z} - \frac{i}{2 k}  \nabla_\perp ^{2} A =  
 i\: \left(\frac{k}{2}n_0^2 r_{33} \tilde E \right)\frac{\partial \phi}{\partial x} A(r,z) \;.
\label{parabolic2}
\end{equation}
Within the same paraxial approximation we drop the longitudinal derivatives in Eq. (\ref{app_pot1}) so that the gradient operator becomes $\nabla_\perp=\hat x \partial/\partial x + \hat y \partial / \partial y.$ Finally introducing the spatial scales $l_\perp$ and $l_\parallel$   and redefining the coordinates and variables through 
$(x,y)/l_\perp \rightarrow (x,y),$
$z/l_\parallel\rightarrow z,$
$\phi (\tilde E/E_{\rm ext})(1/l_\perp)\rightarrow\phi,$
gives the normalized set
\begin{eqnarray}
&&\frac{\partial A}{\partial  z} - i \nabla_\perp ^{2} A =  
 i\: \frac{\partial \phi}{\partial x} A,
\label{parabolic3}
\\
 &&\tau \frac{\partial}{\partial t}\left[\nabla_\perp\cdot\left( \hat \epsilon_n \nabla_\perp\phi \right) \right]+
\nabla_\perp^2\phi+\nabla_\perp\phi\cdot\nabla_\perp\ln\bar I
\nonumber\\
&&= \frac{\partial}{\partial x}\ln\bar I
+\frac{E_{\rm ph}}{E_{\rm ext} \bar I}\frac{\partial I}{\partial x}
+ \frac{k_B T}{eE_{\rm ext} l_\perp} \frac{\nabla_\perp^2 \bar I}{\bar I}
\label{app_pot2}
\end{eqnarray}
where $\tau = t_0/\bar I,$ 
$\bar I = (1+I)/[1- \xi \nabla_\perp\cdot(\hat \epsilon_n \nabla_\perp \phi)],$ 
$\xi=E_{\rm ext}/(\tilde E l_\perp k_D),$ 
$l_\parallel=2/(kn_0^2 r_{33}E_{\rm ext}),$ and  $l_\perp=\sqrt{l_\parallel/(2k)}.$
 
We can calculate characteristic values for the theoretical parameters for the SBN crystal used in the experimental work. 
Using data from Ref. \cite{feinbergsbn} we have
$\epsilon_c=880,$ $\epsilon_y=470,$ and $N_A\simeq 10^{16} ~\rm cm^{-3}$
so that the Debye length is $2\pi/k_D=2.2 ~\mu\rm m,$ and the characteristic 
internal field is $\tilde E = 730 ~\rm V/cm.$
The optical parameters are  $n_0=2.3,$ $\lambda=.532~ \mu\rm m,$ while the electro-optic 
coefficient in our crystal was measured to be $r_{33}\sim 180~\rm pm/V.$ 
With a typical applied field of $E_{\rm ext}=2.2~\rm kV/cm$ 
we have $l_\parallel=350~\mu\rm m,$
$l_\perp=2.5~\mu\rm m,$ $\alpha=0.046,$ and $\xi=0.42.$

\end{document}